# Optimized norm-conserving Vanderbilt pseudopotentials

D. R. Hamann

Department of Physics and Astronomy, Rutgers University, Piscataway, NJ 08854-8019
and Mat-Sim Research LLC, P. O. Box 742, Murray Hill, NJ, 07974

**ABSTRACT**

Fully-nonlocal two-projector norm-conserving pseudopotentials are shown to be compatible with a systematic approach to the optimization of convergence with the size of the plane-wave basis. A new formulation of the optimization is developed, including the ability to apply it to positive-energy atomic scattering states, and to enforce greater continuity in the pseudopotential. The generalization of norm-conservation to multiple projectors is reviewed and recast for the present purposes. Comparisons among the results of all-electron and one- and two-projector norm-conserving pseudopotential calculations of lattice constants and bulk moduli are made for a group of solids chosen to represent a variety of types of bonding and a sampling of the periodic table.

## I. Introduction

While the subject of pseudopotential generation is generally considered to be mature, recent concerns with inaccuracies of tabulated sets of potentials in the context of high-throughput material searches[1,2] indicate that room remains for improvement. Since the introduction of norm-conserving pseudopotentials,[3,4] which in combination with density functional theory[5] paved the way for *ab-initio* calculations of many properties of solids, two main thrusts have driven their improvement. One of these is computational efficiency, and the other is accuracy. Both of these issues can be addressed in other ways. The ultrasoft-pseudopotential method[6] and the related projector-augmented-wave method[7] do so, but at the expense of creating more complex representations of the quantities involved in electronic structure calculations than the simple plane-wave representation of norm-conserving pseudopotentials (NCPPs). While both are routinely used for ground-state energy and structural-relaxation calculations, the implementation of more advanced calculations such as density-functional perturbation theory (DFPT)[8] or many-body perturbation theory[9] becomes vastly more complex than that required with NCPPs. Thus there remains ample motivation to seek further improvements of NCPPs.

Of the two areas for improvement, computational efficiency has received the greater share of attention. The original NCPPs were semi-local, that is, each angular momentum component $\ell$ of a wave function about an atom was acted upon by a different local radial potential. A large step forward in computational efficiency was the transformation of these NCPPs to a local radial potential and a set of separable non-local projectors, one for each of several angular momenta.[10] This "Kleinman Bylander" (KB) approach greatly reduced the computational cost of the Hamiltonial matrix in the plane-wave representation and expedited efficient wave-function evolution methods[11] that dominate electronic structure calculations today. The key properties of semi-local NCPPs were preserved, namely the reproduction of all-electron eigenvalues and integrated total charge inside the core radii $r_{c\ell}$ and the related agreement of the first energy derivatives of the logarithmic derivatives of the outwardly-integrated radial Schrödinger equation at $r_{c\ell}$. At energies further removed from the eigenvalue, however, the log derivatives and hence scattering properties of the all-electron, semi-local, and KB potentials all differ.

The other aspect of computational efficiency which received attention was the rate at which electronic structure results for solids converged with respect to the size of the plane-wave basis. The general prescription for generating NCPPs consists of constructing a node-free pseudo wave function which matches the all-electron wave function to some desired degree of continuity at $r_{c\ell}$ and inverting the radial Schrödinger equation. Several studies analyzed the convergence of the Fourier transform of the semi-local potentials and proposed particular functional forms for the pseudo wave functions that were found to optimize this convergence.[12,13] These approaches gave "one size fits all" prescriptions. A more flexible approach was introduced by Rappe et al., who expanded each pseudo wave function as a linear combination of basis functions and minimized the error of the Fourier-space calculation of its kinetic energy caused by the truncation of this calculation at a cutoff wave vector $q_c$ while satisfying the usual NCPP conditions.[14] Dubbed "optimized pseudopotentials," this method is implemented in the OPIUM open-source code.[15] A portion of this study was motivated by the author's desire to overcome some limitations of that code, and a reformulation of the underlying formalism which couples seamlessly to the accuracy issue to be discussed next is presented in Sec. II.

The accuracy with which *ab-initio* pseudopotentials can predict physical properties of solids is fundamentally limited by that of density functional theory. The ensuing discussion is confined to the ability of NCPPs to reproduce all-electron results. A number of different issues influence accuracy. One is the fact that the ground-state configuration of an atom, usually the best reference in the author's experience, could not be used to generate NCPPs for all the desired angular momenta. A best compromise solution was to use ionized configurations and perhaps fractional occupation of some orbitals to obtain all the required bound states.[4] This limitation was overcome when it was shown that positive-energy scattering states could be used to construct NCPPs.[16] Unfortunately, the Rappe et al. optimization procedure[14] cannot be applied to scattering states because the kinetic energy truncation error cannot be defined. Sec. II also addresses this problem, introducing a soft "barrier" potential beyond $r_{c\ell}$ to create a decaying tail for the reference all-electron function. The scattering properties of NCPPs created in this manner behave essentially identically to those calculated from scattering pseudo wave functions formed using other prescriptions.

A key accuracy issue is the energy range over which an NCPP can reproduce the scattering properties of the all-electron potential. It was observed early on that smaller $r_{c\ell}$s lead to improved agreement, but at the expense of poorer plane wave convergence.[4] Extending the NCPP conditions to require the matching of higher energy derivatives of the radial log derivatives at $r_{c\ell}$ was shown to yield improved agreement,[17] but this has not been widely pursued. It is widely recognized that the choice of local potential in the KB construction changes the scattering properties. A means for adjusting the local potential by adding step functions to one of the semi-local potentials[18] is included in the OPIUM package,[15] and while the steps cancel exactly with the KB projectors at the reference energies, the overall effects on plane-wave convergence may be a cause for concern. In general, there is no systematic prescription for improving the overall scattering properties of a KB NCPP by local-potential adjustment, although trial and error may yield improved results in some cases. A systematic means for improving the scattering properties of fully nonlocal NCPPs was introduced by Blöchl, and involved the introduction of additional separable projectors rather than adjustments of the local potential.[19] He demonstrated that a second projector could greatly improve the agreement of the scattering properties of the semi-local and fully-nonlocal potentials over a range of energies, which is often useful but does not necessarily give significantly improved accuracy compared to all-electron results.

The accuracy focus of the present work is based on Ref. 6, in which Vanderbilt introduced the popular ultrasoft pseudopotentials. It has been widely overlooked that in passing towards the ultrasoft potentials, he gave a prescription for a multiple-projector NCPP which could match scattering properties and norm-conservation to all-electron results at several energies. One proof-of-principal paper[20] was published a few years after Ref. 6, but was not pursued.[21] In Sec. III we review this formalism with revisions appropriate to an NCPP end product, and show how it may be incorporated into our formulation of residual kinetic energy convergence optimization. Combining the terminologies of Ref.14 and Ref.20, we denote these as "optimized norm-conserving Vanderbilt pseudopotentials," (henceforth "OV").

Sec. III also presents comparisons of the KB and OV potentials for the scattering properties of a representative atom. These improvements are found to be especially important in cases where shallow core states are treated explicitly. Convergence is compared for KB and OV, and the manner in which the residual kinetic energy correlates with the total energy convergence for solids[14] is demonstrated. Sec. IV compares OV over KB in results to all-electron results for a selection of solids with ionic, covalent, and metallic bonding incorporating atoms from a variety of positions in the periodic table.

## II. Kinetic energy truncation error optimization revisited

The actual implementation of the optimization principle of Rappe *et al.* is very briefly sketched in the original publication.[14] A careful study of the OPIUM source code[15] shows that that implementation, also undocumented, differs in a number of ways from the initial brief description.[22] The independent approach followed in this work organizes the process in a more transparent manner, and allows for easy extension of optimization to a second Vanderbilt projector. Therefore, we will outline our approach in some detail.

We begin by introducing a generalization of the residual kinetic energy for angular momentum $\ell$ and an operator shorthand notation,

$$E^{\mathrm{r}}_{\ell,ij}(q_{\mathrm{c}}) = \int_{q_{\mathrm{c}}}^{\infty} \varphi_{\ell i}(q)\varphi_{\ell j}(q)q^4 dq \equiv \langle \varphi_{\ell i} | \hat{E}^{\mathrm{r}}(q_{\mathrm{c}}) | \varphi_{\ell j} \rangle \qquad (1)$$

where the Fourier transform

$$\varphi_{\ell i}(q) = 4\pi \int_0^{\infty} j_\ell(qr)\varphi_{\ell i}(r)r^2 dr \qquad (2)$$

is that of a pseudo wave function or a component of a pseudo wave function $\varphi_{\ell i}(r)$. With the exception of references to spherical Bessel functions $j_\ell$, the angular momentum subscript $\ell$ will be omitted below, and it will be assumed that we are working with a single $\ell$ throughout. A diagonal element $i = j$ of Eq.(1) is equivalent to the Rappe *et al.* definition.[14]

Our approach is organized as a hierarchy of radial basis functions which will be denoted as $\xi_i$ and distinguished by various superscripts. The initial set is simply a set of $N$ spherical Bessel functions

$$\xi_i^{\mathrm{B}} = j_\ell(q_i r), r \leq r_c \, ; \, \xi_i^{\mathrm{B}} = 0, r > r_c , \qquad (3)$$

and the choice of wave vectors $q_i$ will be discussed below. The next basis set is the orthonormalized version of $\xi_i^B$,

$$\xi_i^O = \sum_{j=1}^{N} (S^{-1/2})_{ij} \xi_j^B \; ; \; S_{ij} = \langle \xi_i^B | \xi_j^B \rangle. \tag{4}$$

Next, we consider the constraints to be satisfied by the pseudo wave function. In both the original paper and the OPIUM code, continuity of value, slope, and second derivative were required at $r_c$.[14,15] This enforced continuity of value for the semi-local pseudpotential obtained by inverting the Schrödinger equation, but permitted slope discontinuities. These caused us some concern, especially for applications like the calculation of elastic constants via DFPT, where two derivatives of the pseudopotential must be computed.[23] Denoting the final pseudo wave function simply as $\varphi$ and the reference all-electron wave function as $\psi$, we have

$$\varphi = \sum_{i=1}^{N} z_i \xi_i^O, r \le r_c \; ; \; \varphi = \psi, r > r_c, \tag{5}$$

Generalizing the number of continuity constraints at $r_c$ from 3 to M, ie.

$$\frac{d^n \varphi}{dr^n} = \frac{d^n \psi}{dr^n} \equiv d_{n+1}, n = 0, M-1, \tag{6}$$

sets the requirement that the coefficients $z_i$ satisfy the set of M linear equations

$$\sum_{j=1}^{N} C_{ij} z_j = d_i, i = 1, M \; ; \; C_{ij} = \left. \frac{d^{i-1} \xi_j^O}{dr^{i-1}} \right|_{r=r_c}. \tag{7}$$

Proceeding by the standard singular-value decomposition of the $M \times N$ **C** matrix, $\mathbf{C} = \mathbf{U\Sigma V}^T$, we are led to our next set of basis functions. Columns $M+1, \ldots, N$ of **V** correspond to zero singular values of **C**, spanning its null space and yielding our set of null basis functions

$$\xi_i^N = \sum_{j=1}^{N} V_{j,M+i} \xi_j^O, i = 1, N-M. \tag{8}$$

The unique set of coefficients

$$z_{0i} = \sum_{j,k=1}^{M} V_{ij} (\Sigma_{jj})^{-1} U_{jk}^T d_k, i = 1, N \tag{9}$$

defines a component of the desired final pseudo wave function which satisfies all the matching conditions at $r_c$,

$$\varphi_0 = \sum_{i=1}^{N} z_{0i} \xi_i^O \; r \le r_c \; ; \; \varphi = \psi, r > r_c. \tag{10}$$

The previous step influenced the choice of wave vectors $q_i$ defining the $\xi_i^B$ basis set. Our initial choice was to emulate OPIUM, where they are all chosen to match the log derivative of $\psi$ at $r_c$.[15] Since $j_\ell$ are solutions of the spherical wave equation and this requirement imposes homogeneous Robin boundary conditions on $[0, r_c]$, these are eigenfunctions which form an orthogonal set. This choice led to only *M*-1 non-zero singular values of C, indicating an unanticipated linear dependency. In particular, we could not satisfy the 3$^{rd}$ derivative constraint. It became clear that the spherical wave equation imposes relationships among the derivatives of the $j_\ell$. While several alternative choices were satisfactory, all at the expense of the orthonormalization step, the simple expedient of choosing $q_2, q_4, \ldots, q_N$ to match log derivatives and setting $q_1 = q_2/2$ and $q_3 = (q_2 + q_4)/2$ proved very robust.

The members of the basis set $\xi_i^N$ are orthonormal, orthogonal to $\varphi_0$, and have zero value and $M$-1 derivatives at $r_c$. The residual energy to be minimized can now be expressed as

$$E^r = \langle \varphi_0 | \hat{E}^r | \varphi_0 \rangle + 2 \sum_{i=1}^{N-M} \langle \xi_i^N | \hat{E}^r | \varphi_0 \rangle y_i + \sum_{i,j=1}^{N-M} \langle \xi_i^N | \hat{E}^r | \xi_j^N \rangle y_i y_j \quad (11)$$

subject to the norm-conservation constraint

$$\sum_{i=1}^{N-M} y_i^2 = \langle \psi | \psi \rangle_{r_c} - \sum_{i=1}^{N} z_{0i}^2 \equiv D_{\text{norm}}, \quad (12)$$

where $D_{\text{norm}}$ is the "norm deficit" of $\varphi_0$ with respect to the $\psi$ norm on $[0, r_c]$. While Eqs.(11) and (12) constitute a quadratic form to be minimized subject to quadratic constraints, a problem which can conventionally treated by Lagrange multipliers, this did not turn out to be a robust procedure.

Instead, we proceed to our final (promise!) set of "residual" basis functions, $\xi_i^R$, which are formed from linear combinations of the $\xi_i^N$ based on the eigenvectors of the $E_{ij}^r$ matrix in Eq. (11). The corresponding eigenvalues $e_i$ span a very large dynamic range ~$10^6$-$10^8$, which both yields informative insights into the optimization process and suggests an ad hoc minimization procedure which has proven to be very robust. The final pseudo wave function is now

$$\varphi = \varphi_0 + \sum_{i=1}^{N-M} x_i \xi_i^R, \quad (13)$$

and the residual energy is

$$E^r = E_{00}^r + \sum_{i=1}^{N-M} \left( 2 f_i x_i + e_i x_i^2 \right), \quad (14)$$

where the "force" terms $f_i$ are computed from the $E_{i0}^r$ in Eq. (11) using the $E_{ij}^r$ eigenvectors. The norm constraint is Eq. (12) with $y_i$ replaced by $x_i$.

The ad hoc procedure consists of solving the constraint equation for $x_1$ where $e_1$ is the smallest eigenvalue,

$$x_1 = s \left[ D_{\text{norm}} - \sum_{i=2}^{N-M} x_i \right]^{1/2}, \quad (15)$$

and where the sign $s = \pm 1$. The $N - M - 1$ dimensional hypersphere within which the argument of the square root is positive[24] is then searched on a coarse **x** grid, and $E^r$ is evaluated using Eq.(14), trying each choice for *s*. The location of the minimum on this grid is found, along with the corresponding value of *s*. The values of $x_2, ..., x_M$ are then refined iteratively by setting derivatives of Eq. (14) to zero,

$$x_i = -f_i/(e_i + e_1 - sf_1/2x_1), \qquad (16)$$

and recalculating $x_1$ using Eq.(15). This iteration generally converges quickly. As the large dynamic range of the $e_i$ suggests, the law of diminishing returns sets in quite rapidly, with minimal improvements beyond $N = M + 3$ or $N = M + 4$. As a result of this and the simple form of Eq. (14), the dimension of the hypersphere to be searched is small and the energy evaluation trivial.

The optimization procedure outlined above is based on a particular choice of the cutoff wave vector $q_c$, implicit in Eq. (11). After obtaining the $x_i$ coefficients, however, one can transform back to the $y_i$ coefficients of the $\xi_i^N$ basis set and use Eq. (11) to evaluate the function $E^r(q)$ for a range of *q*s. This provides a measure of truncation error per electron which we will see in Sec. III correlates quite well with the actual convergence behavior of the total energy in plane-wave calculations, as claimed in Ref. 14. The optimum choice of $q_c$ is guided by experience, and is typically inversely proportional to $r_c$. For too small a $q_c$, the (typically exponential) convergence of $E^r(q)$ will flatten off for *q*s larger than $q_c$, while for too large a value, useful convergence is unnecessarily sacrificed.

The above formalism may be applied to effectively optimize positive-energy scattering states if the infinite-range oscillatory tail of these states is replaced by a smoothly decaying tail beyond $r_c$. An effective method of achieving this is to add a smooth "barrier" potential to the all-electron potential so that a bound eigenfunction will exist at the desired energy. A satisfactory form is

$$V_{\text{AEB}}(r) = V_{\text{AE}}(r) + v_\infty \theta(x) x^3/(1+x^3) \; ; \; x = (r - r_c)/r_b, \qquad (17)$$

where $\theta$ is the unit step function and the coefficients $v_\infty$ and $r_b$ determine the height and width of the barrier. Since the value and two derivatives of the barrier function vanish at $r_c$, 4 derivatives of the all-electron eigenfunction are continuous. While this choice is ad hoc and the shape of the tail effects the values of the terms in Eq.(11), the optimized pseudo wave function is quite insensitive to the barrier parameters for sensible choices. The log derivative at $r_c$ is determined by the energy alone, and is identical to that which would have been obtained with the original scattering-state method.[16] The barrier $\psi$ should have one more node than the highest-lying core state with the same $\ell$, or no nodes if there is none. In the next section, when an additional $\psi$ is required at higher energy, another node should be added.

The barrier method might typically be employed to generate the *d* pseudopotentials for atoms with no valence *d* electrons. In Fig. 1, we illustrate the relations among the all-electron and pseudo wave functions for barrier-bound and scattering Ge *d* states, calculated at an energy of +0.25Ha.

It was remarked in Ref. 13 that while pseudopotentials calculated with the Rappe et al. method[14] do show fast total energy convergence in the solid, they display "strong short-wavelength oscillations." Those authors advanced transferability concerns as a motivation for pursuing their approach to improving convergence.[13] Such oscillations appear for a number of the potentials produced using the sample input data distributed with the OPIUM code.[15] This does not seem to be the case using the optimization algorithm described above. The set of smooth Ge pseudopotentials shown in Fig. 2 is typical of what we find. While we can only speculate on the differences, it is possible that the surprisingly large dynamic range of the $E_{ij}^{\rm r}$ eigenvalues plays a role. OPIUM appears to use a global conjugate-gradient minimization method, which may not treat what are in effect the $x_i$ coefficients of the less-important $\xi_i^{\rm R}(r)$ in Eq.(13) as accurately as do Eqs.(15) and (16).

The effects of increasing the continuity of the optimized pseudopotentials at $r_c$ are less obvious. Ca *d* semi-local potentials, calculated for an unoccupied shallow bound 3d state, are shown in Fig. 3 with value-only, first-derivative, and second-derivative continuity. They are very similar, and it takes the magnification of the inset to really discern the continuity differences. It is apparent that the optimization procedure tends to suppress discontinuities even when they are not strictly eliminated.

## III. Optimizing Vanderbilt projectors

We will briefly review Vanderbilt's derivation,[6] pausing and introducing an appropriate simplification at the point where the norm-conserving and ultrasoft versions diverge. Several reference all-electron wave functions $\psi_i$ and corresponding pseudo wave functions $\varphi_i$ at energies $\varepsilon_i$ will be considered, all at a given $\ell$ as usual. Let these actually be $r$ times the radial wave functions so that the kinetic energy operator simplifies to $T = [-d^2/dr^2 + \ell(\ell+1)/r^2]/2$ in atomic units. Let us choose a local potential $V_{\rm loc}$ which joins smoothly to $V_{\rm AE}$ at some $r \leq r_c$ but is otherwise arbitrary. Following Ref. 6, we introduce the projectors

$$|\chi_i\rangle = (\varepsilon_i - T - V_{\rm loc})|\varphi_i\rangle. \qquad (18)$$

Note that $\chi_i$ are zero for $r \geq r_c$.

For a single projector $\chi_1$, the non-local potential operator is

$$V_{\rm NL} = \frac{|\chi_1\rangle\langle\chi_1|}{\langle\varphi_1|\chi_1\rangle}, \qquad (19)$$

Which is the usual KB result,[10] although obtained without any reference to the semi-local potential. Vanderbilt's generalized this to the case of more projectors, and we will revise his Eq. (7) for our purposes[25] to

$$V_{\text{NL}} = \sum_{i,j} |\chi_i\rangle (B^{-1})_{ij} \langle \chi_j|, \tag{20}$$

where

$$B_{ij} = \langle \varphi_i | \chi_j \rangle. \tag{21}$$

Now in general, $B_{ij} \neq B_{ji}$, so $V_{\text{NL}}$ would be a non-Hermitian operator. However, after performing integration by parts on the integrals giving $B_{ij}$,

$$B_{ij} = \int_0^{r_c} \varphi_i \left[ \varepsilon_j + \frac{1}{2} \frac{d^2}{dr^2} - \frac{\ell(\ell+1)}{2r^2} - V_{\text{loc}} \right] \varphi_j, \tag{22}$$

and subtracting corresponding expressions for the all-electron $\psi_i$ with $V_{\text{AE}}$, he proved that $B_{ij}$ will be a symmetric matrix if the generalized norm-conservation condition

$$\langle \varphi_i | \varphi_j \rangle_{r_c} = \langle \psi_i | \psi_j \rangle_{r_c} \tag{23}$$

is satisfied, where the $r_c$ subscripts indicate that the overlap is to be computed on $[0, r_c]$.

Vanderbilt went on to show that for the ultrasoft case, ie. allowing Eq.(23) to be violated and compensating appropriately, the energy derivatives of the log derivatives of pseudo wave functions calculated from $V_{\text{loc}} + V_{\text{NL}}$ will match those of corresponding all-electron functions at each $\varepsilon_i$.[6] This also applies to the generalized norm-conserving case, where Eq.(23) is satisfied, extending this property of the original semi-local pseudopotentials[3] to several energies.

For purposes of ease of integration with plane-wave codes, we have transformed Eq.(20) one step further, normalizing the $\chi_i$, rescaling $B_{ij}$ appropriately, diagonalizing it, and forming linear combinations of the $\chi_i$ using the resulting eigenvectors. Our final form for the non-local operator is

$$V_{\text{NL}} = \sum_i |\tilde{\chi}_i\rangle \frac{1}{\tilde{b}_i} \langle \tilde{\chi}_i|, \tag{24}$$

where $\tilde{b}_i$ are the eigenvalues of the rescaled $B_{ij}$.

In general we prefer to use the scalar-relativistic radial Schrödinger equation[26] for our all-electron calculations, since by including the mass-velocity, Darwin, and other higher-order terms it gives a better description of heavier atoms. Since the kinetic energy is no longer the simple second derivative, the integration-by-parts subtractions of Eq.(22) and its all-electron analogue no longer cancel, so the exact symmetry of $B_{ij}$ is not ensured. In practice, we find that the asymmetry is $\sim 10^{-4}$ to $10^{-5}$ for both light and heavy atoms, so we simply symmetrize $B_{ij}$ and proceed. This manifests itself in disagreements of comparable magnitude in comparisons of quantities such as eigenvalues and norms computed with the final OV potentials, which are typically correct to $\sim 10^{-8}$ when non-relativistic all-electron calculations are employed. While they have not yet been implemented, we expect similar behavior for fully-relativistic calculations employing the Dirac equation.

Incorporating the norm-conserving Vanderbilt construction into the residual energy optimization framework,[14] we will restrict our attention to two projectors. The reference $\psi_1$ and $\psi_2$ for a given $\ell$ might be chosen to be a shallow core and a valence wave function, a valence and a barrier function, or two barrier functions. Typically, we find a spread of ~1Ha between $\varepsilon_1$ and $\varepsilon_2$ works well when the choice is not dictated by the use of two bound functions. It is appropriate for $\psi_2$ to have one more node than $\psi_1$ inside $r_c$. The procedures of Sec. II are followed to construct a nodeless norm-conserving $\varphi_1$.

The key observation in proceeding to the calculation of $\varphi_2$ is that while the diagonal terms in Eq.(23) are quadratic constraints, the off-diagonal term is a linear constraint. Since $\psi_2$ will in general have a different log derivative at $r_c$ than $\psi_1$, we could go all the way back to the beginning of our basis set construction. However, the orthonormal $\xi_i^O$ basis calculated for $\varphi_1$ has proven to be perfectly adequate for $\varphi_2$. The off-diagonal norm constraint can now imposed simply by adding a row to the constraint matrix $C_{ij}$ in Eq.(7) and an element to the "derivatives" vector,

$$C_{M+1,i} = z_i \ ; d_{M+1} = \langle \psi_1 | \psi_2 \rangle_{r_c} \qquad (25)$$

where $z_i$ is the set of $N \xi_i^O$ coefficients in Eq.(5). They are formed as the sum of the $z_{0i}$ coefficients in Eq.(10) and the are the corresponding $\xi_i^O$ coefficients transformed back from the optimized $x_i$ in Eq.(13) using the $E_{ij}^r$ eigenvectors and the null singular vectors of the original $C_{ij}$. The optimization of $\varphi_2$ now proceeds as in Sec.II from Eq.(7) onward, with the quadratic 2,2 normalization constraint of Eq.(23) treated as in Eq.(12)

The new null basis set $\xi_i^N$ for $\varphi_2$ now has one fewer member than that for $\varphi_1$, so in principal $E^r$ cannot be as well optimized. In practice, $\varphi_2$ is either a pseudo valence state which despite its single node is intrinsically "softer" in $q$-space than the corresponding shallow-core $\varphi_1$, or is a scattering state sufficiently higher in energy that it does not enter into the occupied states in the solid with appreciable amplitude. An example of each case is shown in Fig. 4, where the convergence of the total energies of Si and Cu are plotted as functions of plane-wave cutoff energy for KB and OV calculations. Differences are basically negligible in the relevant range of cutoffs. This figure also confirms the manner in which $E^r(q)$ correlates with the actual plane-wave behavior, where we have plotted it for the least-rapidly converging $\ell$ for each material. (We note in passing that the pseudo Cu 3s and 3p states converge more rapidly than the 3d, so there is no significant computational penalty in treating them as valence.)

The improvement in reproducing all-electron (AE) scattering results with KB and OV pseudopotentials is illustrated in Fig. 5 for K. The 3s and 3p shallow core states are treated as valence, and the local potential is a smooth polynomial extrapolation of the AE potential from the minimum $r_c$ to zero. The arc tangents of the log derivatives at $r_c$, which are somewhat analogous to scattering phase shifts are plotted. These are much easier to compare visually than the log derivative themselves. The AE and OV results are identical within the linewidths, while the KB results deviate significantly for the s and d channels. This is consistent with the one example in Ref. 20, and representative of all the OV pseudopotentials used in the tests in the next section. While the log derivative error appears to be quite small at the -0.089 Ha binding energy of the K 4s state, the differences between the OV and KB pseudo wave functions shown in Fig. 6 are significant. The OV result reproduces the AE results perfectly outside $r_c$ by construction.

A problem that must be addressed with any Hamiltonian containing a separable non-local operator like Eq.(24) is the fact that its eigenstates are not necessarily ordered in energy by numbers of nodes. So-called "ghost states" at energies below the nodeless pseudo wave function from which the pseudopotential was generated can invalidate results. An analysis of the KB case gives a straightforward prescription to test for this possibility.[27] This does not generalize to the multi-projector case, but we can test a potential by scanning the log derivative it produces outside $r_c$ over a sufficiently wide range of energies below the lowest desired eigenvalue. A spurious step in a plot such as Fig. 5 signals the occurrence of a ghost. In practice, the second projector of the OV method is very effective at suppressing ghosts compared to KB. In either case, adjustment of the local potential will fix the problem.

## IV. Results for solids

The appropriate tests are to compare all-electron density-functional calculations for solids with pseudopotential calculations. For our reference calculations, we used the open-source ELK code, which employs the full-potential linear-augmented-planc-wave plus local-orbital method.[28] The default parameters appear to yield well-converged results, and are employed for all the calculations except for a few cases in which muffin-tin radii had to be decreased to accommodate short bond lengths. The calculations used here are effectively scalar-relativistic, based on weighted averages of Dirac equation solutions within the muffin tines. The local density approximation was used.[29] Since ELK is not able to optimize lattice parameters directly, we chose cubic materials for all but one case, so that energy vs. volume curves fitted with the Burch-Murnaghan equation of state[30] could easily yield the lattice constant $a$ and the bulk modulus $B_0$. While this equation of state was used throughout for consistency among the quoted results, five other functional forms available within ELK[28] were tested for several cases. The spreads among the results were $\pm 0.001 a_B$ for $a$ and $\pm 1-2\%$ for $B_0$, with Burch-Murnaghan typically falling at the center of the distribution.

Plane-wave calculations were carried out using the ABINIT code.[31,32] Full structural optimization was carried out via force and stress minimization, and the bulk modulus was determined from elastic constants calculated using DFPT.[23] Well-converged Brillouin-zone samples, Fermi smoothing of band occupations for metals, etc. were kept consistent between the AE and pseudopotential calculations. Plane wave convergence was tested, and the lattice constants presented in Table I are all converged to ~0.1% and the bulk moduli to ~1% or better at the stated cutoff energies.

The test cases were chosen to represent a variety of types of bonding and to involve atoms which give a representative, if coarse, sampling of the periodic table. Most atoms were used in two and sometimes three solids, always represented by the same pseudopotentials. All the cases we tested have been included, whether or not there was significant improvement in agreement using OV potentials. All pseudopotentials were based on the atomic ground-state configuration. All parameters for each element were identical for the KB and OV potentials. Projectors for s, p, and d were included for all but first-row atoms, with f projectors for two atoms. $V_{loc}$ was a smooth polynomial extrapolation of $V_{AE}$ in all cases. Semi-core electrons mentioned explicitly were treated as valence in the calculations. The potentials and test solids are discussed in tabular order below.

The K calculations included 3s, 3p, and 4s. Two K potentials were used. The initial KB results for bcc K metal showed sufficient errors that the K* potential was tried using smaller $r_c$s, which are generally found to improve results (if at the expense of convergence). The OV results are in excellent agreement with AE, and identical for both sets of parameters, while the KB results bracket AE, with somewhat better agreement for K*. Moving to the ionic insulator KCl, where Cl included only the outer 3s and 3p, the OV results are once again in excellent agreement with AE and identical for K and K*. The KB results for both K and K* are in qualitatively worse agreement than for K metal, with no apparent correlation between the lattice-constant and bulk-modulus errors.

To test a different structure and give K* another chance, we chose KBaN, a half-Heusler-structure insulator, not yet known experimentally but recently proposed as a promising piezoelectric.[33] N 2s and 2p, and Ba 4d and 6s electrons are included. AE and OV results are in excellent agreement, while KB show substantial errors. Ba as an elemental bcc metal provides another test for its potentials. For OV, $a$ is in excellent agreement and $B_0$ off by +3%, while corresponding errors for KB are +5% and +9%.

Providing another test for N and introducing a 4d transition metal, we chose the metallic rock-salt compound ZrN. Zr 4s, 4p, 4d, and 5s electrons were included. In general, we found that transition-metal d electrons limited convergence, and that including semi-cores in the same shell added little computational effort. We found excellent agreement for OV, and substantial errors for KB.

La has a bound but unoccupied 4f state which should influence its bonding in a solid, so we added rock-salt LaN as a test. La 5s, 5p, 5d and 6s electrons were included. Although anticipated to be an insulator, it was semi-metallic within the local density approximation. The $E^r$ analysis of the optimized 4f pseudo wave function suggested a 30

Ha cutoff, which was apparently necessary. OV results were excellent, but once again KB showed substantial errors.

Moving to the center of the periodic table, Si showed excellent results for both OV and KB at a modest cutoff, providing another example of an old adage of the electronic-structure community: "Anything works for Si." Only Si 3s and 3p electrons were included, along a non-linear core correction charge[34] in polynomial form.[35] To provide more of a challenge, we studied $SiO_2$ in an artificial cubic $Fd\bar{3}m$ structure once mistakenly thought to be that of $\beta$ crystobalite. It is best described as an expanded diamond lattice of Si with O inserted midway between each Si neighbor pair, ie. with $180^0$ Si-O-Si bond angles. Optimizing this structure gives the essentially standard Si-O bond length of 1.6Å, and so this hypothetical material should be reasonably representative of real $SiO_2$ bonding. In this case, too, both OV and KB are in good agreement with AE. To push this case one step further, we took advantage of the O-required cutoff and introduced the Si* pseudopotential, with the rather deep 2s and 2p core states treated as valence. The optimization procedure was very effective, and the OV results remained extremely well converged at 30 Ha ($4 \times 10^{-5}$ $a_B$ and 0.1 GPa compared to 40Ha). Unfortunately, Si* with KB did not even bind the solid, the energy being a monotonically decreasing function of $a$ for $11 \leq a \leq 20$ $a_B$. The corresponding Tab. I entries are labeled "Not Available" (NA).

A third Si-based material, the metallic compound $CaSi_2$, was included because it has the unusual property of showing significant occupation of a Ca 3d state, which is weakly bound and unoccupied in the atom.[36] Including the 3s, 3p, and 4s electrons for Ca, it is the 3d that controls convergence in the solid, and makes this system an interesting test for both convergence optimization and the effect of the second projector. Among several polymorphs, we used the trigonal rhombohedral "tr3" structure,[37] space group $R\bar{3}m$, with one formula unit per primitive cell. Qualitatively, buckled Si double layers similar to (111) double layers in the Si diamond structure are separated by intercalated Ca atoms. Structural relaxation using ELK was accomplished using a mesh of ~30 $a$ and $c$ lattice constants, relaxing the single internal coordinate, and fitting the resulting energies with a cubic polynomial in $a$ and $c$. The $B_0$ calculation using DFPT within ABINIT was supplemented by a relaxation correction using DFPT internal strain and interatomic force constants.[23] For this system, the OV results are in excellent agreement with AE, but the KB results show substantial errors, with the differences presumably arising mainly from the Ca. (A second $CaSi_2$ row has been added to Table I for the $c$ lattice constant.)

The cubic perovskite structure of $SrTiO_3$ was chosen to have more typical 3d hybridization in an insulator. Sr 4s, 4p, and 5s were included, with 3s, 3p, 3d, and 4s for Ti. Once again, the Sr and Ti semi-cores did not limit convergence. The OV results are in excellent agreement with AE. KB shows a moderate -4% error for $a$, and a significantly larger error for $B_0$. Dropping the Ti, rock-salt SrO shows comparable levels of agreement and disagreement for OV and KB.

The rock-salt metal BiSe was included to have a scattering-state f projector above the filled Bi 4f core level, with 5d, 6s, and 6p included as valence. Se included only 4s and 4p. For this compound, both OV and KB gave excellent agreement with AE.

Elemental fcc Cu was included primarily for historical reasons, its 3d potential being the first published optimized pseudopotential.[14] The rather prominent slope discontinuity seen at $r_c$ in Fig. 2 of Ref. 14 initially motivated part of the present work. With the optimization approach described in Sec. II, however, the unconstrained slope discontinuity for Cu 3d was even less apparent that that shown in our Fig. 3 for Ca. The results for the solid, with 3s, 3p, 3d, and 4s show excellent AE-OV agreement, with substantial errors for KB. The 3d sets the convergence behavior, shown in Fig. 4.

## V. Discussion and conclusions

The main conclusion of the work described above is that the neglected Vanderbilt approach to norm-conserving multi-projector pseudopotentials[6] can be used in the context of systematic convergence optimization,[14] and can serve as a competitive choice for accuracy and computational efficiency compared to ultrasoft[6] and projector-augmented-wave potentials.[7] Some trends are discernable among the results for the 12 solids studied as test cases. The outermost core electrons were treated as valence for many of the atoms in these tests. This is widely known to be particularly important to obtain accurate results for group 1 and group 2 elements, and can also be important for transition-metal elements. The greatest differences between OV and KB results occurred in cases with these cores. With optimization, the plane-wave cutoff requirements with core states remained relatively modest.

The second observation to be drawn is that the use of the ground-state configurations of the atoms to generate the OV potentials gives comparably good results in elemental, covalent, and ionic solids. In fact, the test systems involving Si included strictly covalent diamond Si, cationic Si in $SiO_2$, and at least in electronegativity terms, anionic Si in $CaSi_2$.

A third comment is that the KB results for any given system can undoubtedly be improved by further adjustments of the local potential choice and core radii. While in the course of this research, some of these parameters were changed to eliminate ghost states or improve very bad KB results, the OV results and their agreement with AE were essentially unchanged. Examining the log derivative plots gives some guidance to improving KB, but there is no systematic approach. The OPIUM code[15] provides the ability to compare all-electron and pseudopotential energy differences between the reference atomic configuration and other configurations. This capability is incorporated into the ONCVPSP code developed for this research, and all test configurations involved one- and two-electron ionized states since the neutral ground state was the reference. For the atoms used in the test cases, the rms excitation error was 0.012 Ha for KB vs. 0.003 Ha for OV. However, atom-by-atom, the correlation between these results and results for solids was at best difficult to discern.

A final point to be made is that in no case among the 14 atoms in the tests were the parameters used in constructing the OV potentials adjusted to "improve" agreement with the AE results. The agreement was nearly always within the range of the spread of the equation-of-state fits, and usually better. Some experimentation with the optimization parameters $q_c$ and $N$ as well as the projector energies $\varepsilon_1$ and $\varepsilon_2$ (where not fixed by bound states) was done to improve convergence and balance it among the $\ell$s, but this was all evaluated within the confines of the pseudopotential generation code, with no reference to results for solids. With the exception of the "Si* experiment," decisions on treating core states as valence were made in advance of any comparisons.

While all the calculations reported here were done using the local density approximation,[29] ELK AE results and OV pseudopotential results were also compared using the PBE generalized-gradient functional[38] for several of the test systems. Agreement was comparable. However, when a pseudopotential generated with PBE was (unintentionally) used to compare local-density AE and OV calculations, differences increased noticeably.

The overall conclusion of this research is that the accuracy of two-projector OV pseudopotentials in calculating the properties of materials is primarily limited by the accuracy of the underlying density functional approximations. The open-source ONCVPSP code is freely available.[39]

**Acknowledgements**

The author would like to acknowledge valuable discussions with D. Vanderbilt, K. M. Rabe, and J. W. Bennett.

Table I. Comparisons of lattice constants and bulk moduli among all-electron, optimized Vanderbilt, and Kleinman-Bylander calculations for the test set of solids. K* and Si* are explained in the text. (a) NA (Not Available) is also explained in the text.

| System | $E_{cut}$(Ha) | Lattice constants ($a_B$) | | | Bulk moduli (GPa) | | |
|---|---|---|---|---|---|---|---|
| | | AE | OV | KB | AE | OV | KB |
| K | 20 | 9.58 | 9.58 | 10.56 | 4.14 | 4.13 | 3.00 |
| K* | 20 | 9.58 | 9.58 | 9.33 | 4.14 | 4.13 | 3.83 |
| KCl | 20 | 11.49 | 11.48 | 14.64 | 23.90 | 24.33 | 8.69 |
| K*Cl | 20 | 11.49 | 11.48 | 11.14 | 23.90 | 24.36 | 31.53 |
| K*BaN | 25 | 12.23 | 12.25 | 12.70 | 44.01 | 44.22 | 34.35 |
| Ba | 20 | 9.11 | 9.09 | 9.57 | 9.22 | 9.47 | 10.01 |
| ZrN | 25 | 8.55 | 8.57 | 9.49 | 280.98 | 280.50 | 154.48 |
| LaN | 30 | 9.84 | 9.83 | 8.24 | 134.84 | 134.67 | 222.44 |
| Si | 10 | 10.21 | 10.21 | 10.20 | 97.11 | 95.80 | 95.30 |
| $SiO_2$ | 30 | 13.90 | 14.07 | 14.09 | 152.68 | 152.69 | 152.88 |
| $Si*O_2$ | 30 | 13.90 | 13.92 | NA[a] | 152.68 | 155.32 | NA[a] |
| $CaSi_2$ | 20 | 7.14 | 7.14 | 6.94 | 68.42 | 67.05 | 88.11 |
| $CaSi_2$ c | | 29.30 | 29.19 | 26.71 | | | |
| $SrTiO_3$ | 30 | 7.27 | 7.31 | 6.95 | 198.85 | 201.51 | 269.03 |
| SrO | 30 | 9.56 | 9.58 | 8.79 | 105.73 | 104.40 | 194.04 |
| BiSe | 20 | 11.38 | 11.40 | 11.40 | 66.03 | 65.81 | 66.45 |
| Cu | 30 | 6.57 | 6.58 | 7.27 | 172.47 | 174.17 | 106.95 |

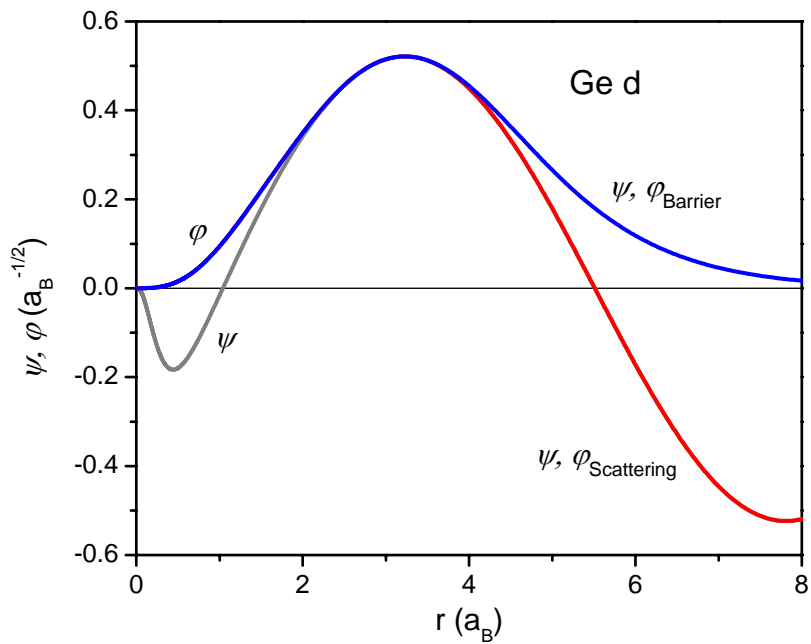

Fig. 1. All-electron ($\psi$) and pseudo ($\varphi$) wave functions for the Ge d scattering state at 0.25 Ha, illustrating the use of a soft barrier to induce a bound-state-like tail which allows residual energy optimization.

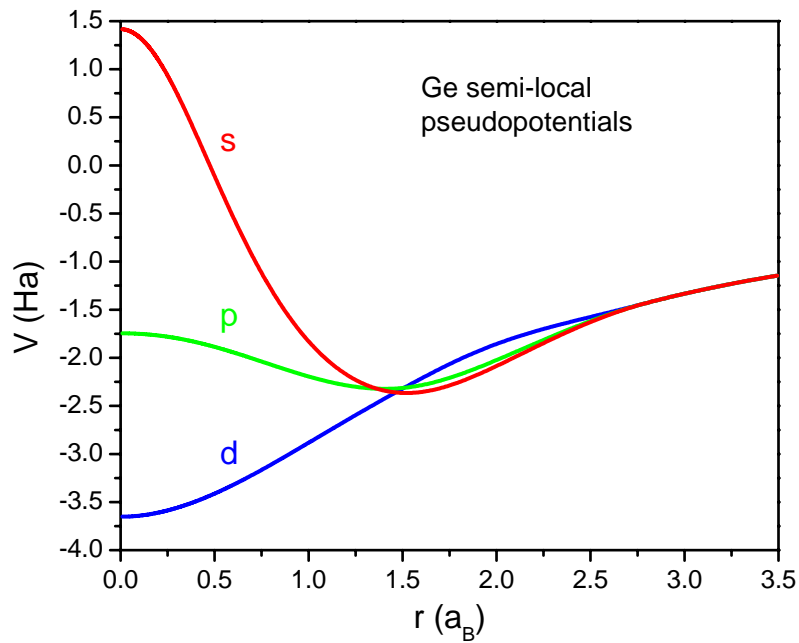

Fig. 2. Ge pseudopotentials illustrating the smooth behavior characteristic of the residual energy optimization approach introduced here.

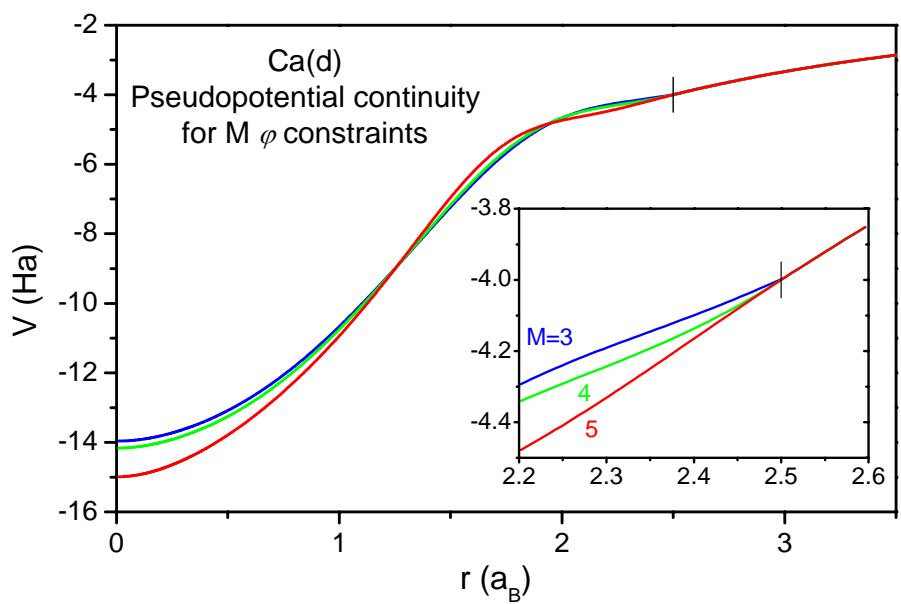

Fig. 3. Illustration of the pseudopotential continuity results from requiring M=3, 4, or 5 value plus derivative continuity constraints on the pseudo wave function, for the weakly-bound unoccupied Ca d state. $r_c = 2.5$ $a_B$ is indicated by the vertical bar.

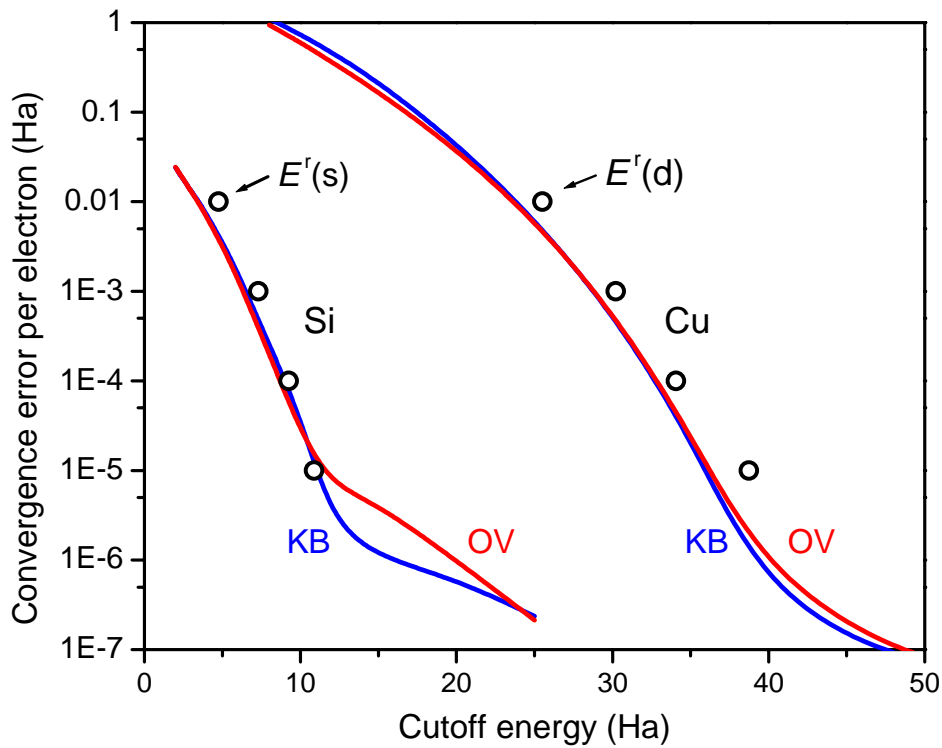

Fig. 4. Convergence of the total energy with plane-wave cutoff for Si and Cu solids, showing the minor effect of the second projector of the OV method compared to KB. The points are residual energies from the first-projector atomic calculations.

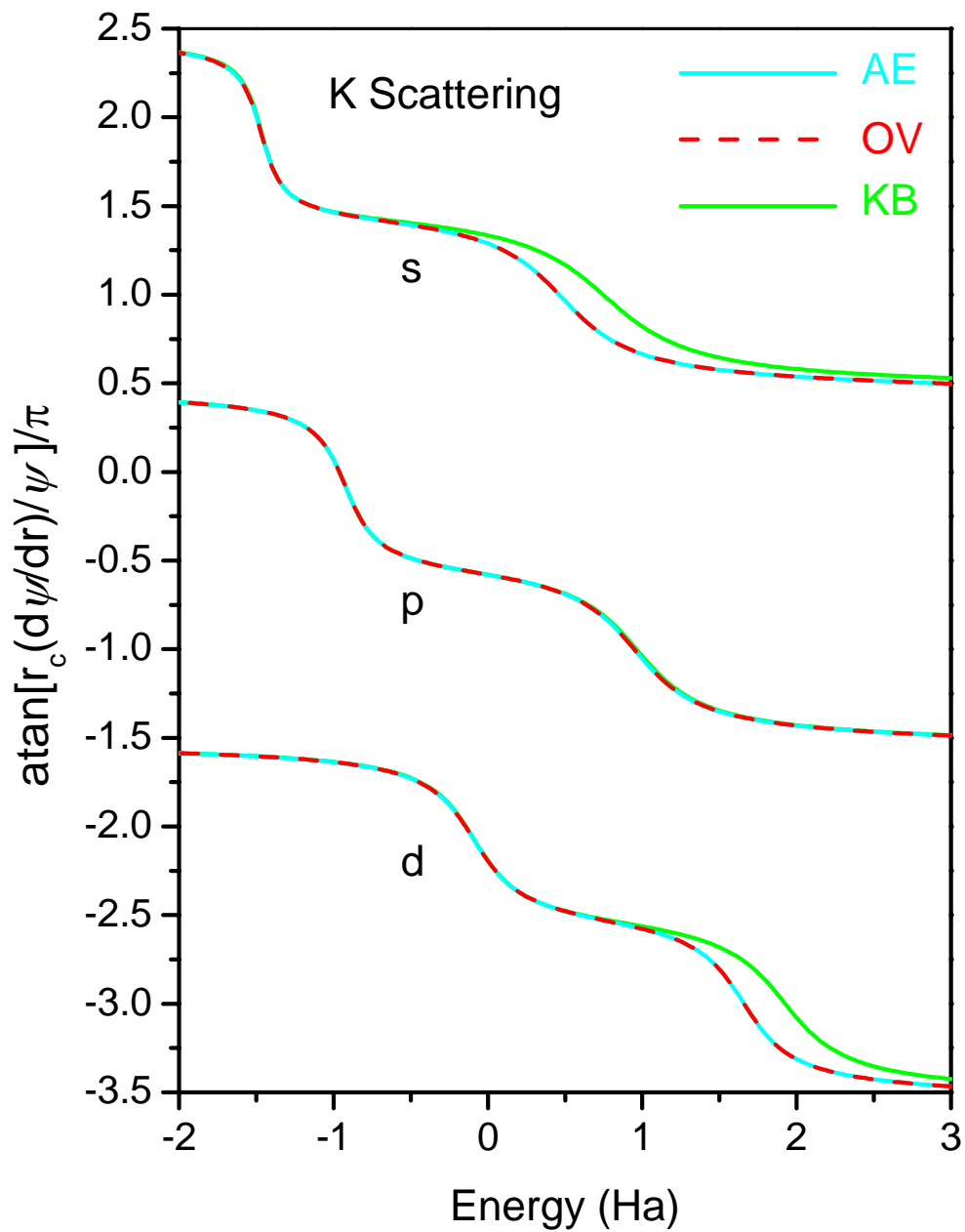

Fig. 5. K log derivatives vs. energy plotted as $\mathrm{atan}[r_c(d\psi/dr)/\psi]/\pi$ at $r_c$. All-electron, OV, and KB are compared.

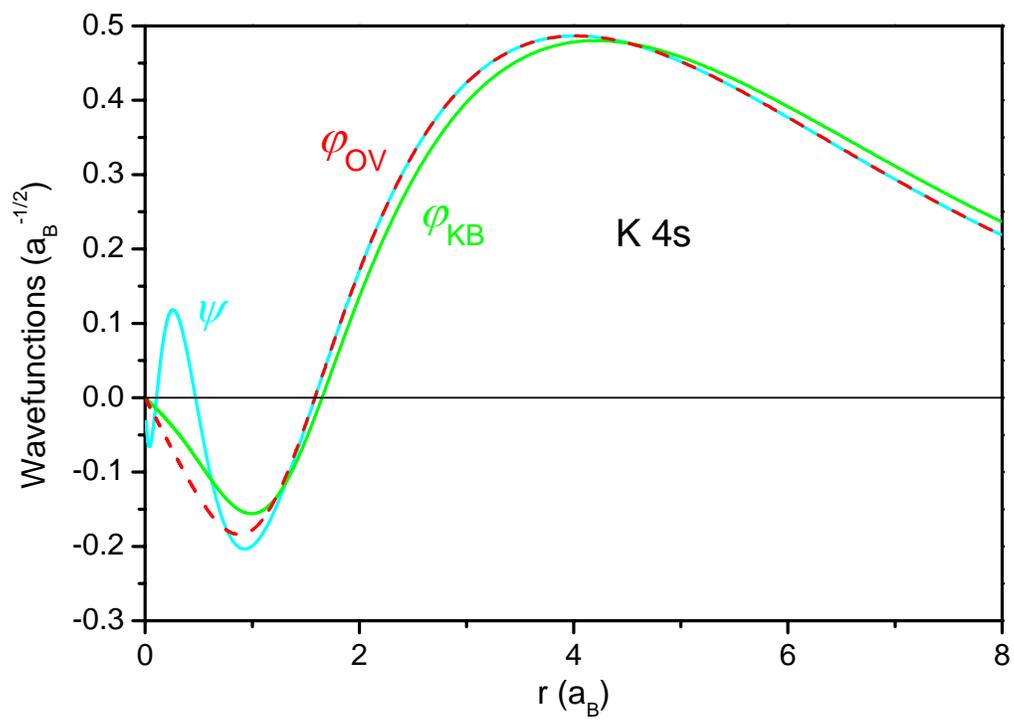

Fig 6. K 4s wave functions computed with the all-electron potential, and KB and OV pseudopotentials.